\documentclass[11pt]{article}
\usepackage{hyperref}
\usepackage{tikz}

\usepackage{amsmath,amsthm,amsfonts,amssymb,amsopn,amscd} 
\DeclareMathOperator{\Ind}{Ind}

\def\a{\alpha}
\def\b{\beta}

\def\e{\varepsilon}

\def\l{\lambda}

\def\r{\rho}

\def\om{\f}
\def\Om{\xi}

\def\gA{{\mathfrak A}}

\def\A{{\cal A}}

\def\M{{\cal M}}
\def\N{{\cal N}}
\def\R{{\cal R}}

\def\H{{\cal H}}
\def\K{{\cal K}}
\def\S{{\cal S}}

\def\f{{\varphi}}
\def\s{{\sigma}}

\def\l{{\lambda}}
\def\x{{{h}}}

\def\PSL{{{\rm PSL}(2,\mathbb R)}}

\def\S2{S^{1(2)}}

\theoremstyle{definition} 

\theoremstyle{remark} 

\newcommand{\ben}{\begin{equation}}
\newcommand{\een}{\end{equation}}

\def\PSL{PSU(1,1)}

\def\SL2{{{\rm SL}(2,\R)}}

\def\PSL2{{{\rm PSL}(2,\Reali)}}

\def\U1{{{\rm V}(1)}}
\def\SU2{{{\rm SV}(2)}}

\def\SU{{{\rm SU}}}

\def\A{{\mathcal A}}

\def\H{{\mathcal H}}

\def\K{{\mathcal K}}

\def\M{{\mathcal M}}
\def\N{{\mathcal N}}

\def\P{{\mathcal P}}

\title{\Huge{The emergence of time 
}}
\author{{\sc Roberto Longo}\thanks{Supported by the ERC Advanced Grant 669240 QUEST ``Quantum Algebraic Structures and Models'', MIUR FARE R16X5RB55W  QUEST-NET and GNAMPA-INdAM. \eject
E-mail: {longo@mat.uniroma2.it}
}
\\
Dipartimento di Matematica,
Universit\`a di Roma Tor Vergata,\\
Via della Ricerca Scientifica, 1, I-00133 Roma, Italy
}

\date{}
\begin{document}

\maketitle

\begin{abstract}
Classically, one could imagine a completely static space, thus without time. As is known, this picture is unconceivable in quantum physics due to vacuum fluctuations. The fundamental difference between the two frameworks is that classical physics is commutative (simultaneous  observables) while quantum physics is intrinsically noncommutative (Heisenberg uncertainty relations). In this sense, we may say that time is generated by noncommutativity; if this statement is correct, we should be able to derive time out of a noncommutative space.

We know that a von Neumann algebra is a noncommutative space. About 50 years ago the Tomita-Takesaki modular theory revealed an intrinsic evolution associated with any given (faithful, normal) state of a von Neumann algebra, so a noncommutative space is intrinsically dynamical. This evolution is characterised by the Kubo-Martin-Schwinger thermal equilibrium condition in quantum statistical mechanics (Haag, Hugenholtz, Winnink), thus modular time is related to temperature. Indeed, positivity of temperature fixes a quantum-thermodynamical arrow of time. 

We shall sketch some aspects of our recent work extending the modular evolution to a quantum operation (completely positive map) level and how this gives a mathematically rigorous understanding of entropy bounds in physics and information theory. A key point is the relation with Jones' index of subfactors. 

In the last part, we outline further recent entropy computations in relativistic quantum field theory models by operator algebraic methods, that can be read also within classical information theory. 
The information contained in a classical wave packet is defined by the modular theory of standard subspaces and related to the quantum null energy inequality. 
\end{abstract}

\newpage

\section{Introduction}

In 1967, Richard Kadison organised the first conference on Operator Algebras at Baton Rouge, an event that set together essentially all the experts in the subject at that time, both from Mathematics and Physics, and put the seed for much of the future developments in the subject.

Due to  a remarkable historical accident, two related breakthroughs were announced at that conference. M. Tomita distributed a (never published) preprint on modular theory \cite{Tom}, that set a profound and fundamental change of view in the subject. R. Haag presented in a talk the characterisation of equilibrium states in Quantum Statistical Mechanics at infinite volume \cite{HHW}, extending the Gibbs condition: the KMS (Kubo-Martin-Schwinger) property. 

During the next year, fervid discussions on modular theory took place, in particular at Kadison's seminar on Operator Algebras at Penn. Eventually, Masamichi Takesaki provided a complete, extended, rigorous formulation of modular theory and showed that the modular group is characterised by the KMS condition \cite{Tak}. Clearly, this was the basis for a new deep interplay between Mathematics and Physics, that is still producing impressive results. 

In this article, we briefly outline a path from these general basic results to some present developments where the interplay between Operator Algebras and Quantum Field Theory is playing a key role.

We refer to \cite{KR} and to \cite{Araki, Haag} for the basics on Operator Algebras and on Local Quantum Field Theory. 

\section{General framework}

\subsection{Space and time}
We begin with general considerations of interdisciplinary nature about space and time, that explains why we expect Operator Algebras and modular theory to play a even more central role in Science in the future. 

From a mathematical point of view, a space is a set $X$ endowed with a structure that put relations among elements (points): for example a distance, a topology, a measure. Time then allows to compare different configurations, in particular it provides a (partial) order. 

Now, in Quantum Physics, space is never static (due to spontaneous creation/annihilation of particle-antiparticle pairs) so, in a sense, time should emerge out of the space structure. As said, one main property of time is that it has an order, the \textit{arrow of time}, that we can observe both macroscopically in classical physics and microscopically in quantum physics. 

Classically, the arrow shows up by the second principle of thermodynamics, the entropy of an irreversible process never increases. On the other hand, in quantum mechanics, a wave function is set in a pure state after observation, an irreversible process that is unidirectional in time, so:
\[
\text{thermodynamics}\rightarrow \text{positive entropy}
\]
\[
\text{quantum mechanics}\rightarrow \text{collapse of the wave function}
\]
A known question is to relate the two arrows of time. An optimal answer would be to derive one arrow from the other. A less stringent, still interesting question is the following: 
\[
\text{is there a general frame to encompass both arrows?}
\]
Of course, we keep in mind that time is a relative concept as we have learnt from Einstein's Theory of Relativity. 

\subsection{Quantum Physics and Noncommutativity}
As is  known, the fundamental difference novelty of quantum physics with respect to classical physics is {\it noncommutativity}: observables are represented by selfadjoint operators on a Hilbert space $\H$, expectation values are given by vector states.  Heisenberg commutation relation between position and momentum  operators
\[
PQ - QP = i I
\]
shows that these two observables, in particular, do not commute.

By the spectral theorem, a selfadjoint operator is determined by its bounded continuous functional calculus, so we may restrict our attention to bounded selfadjoint operators.
A \textit{quantum system} is then given by a unital, selfadjoint algebra of bounded linear operator $\M$ on $\H$. If we fix our representation for the moment, we may assume that $\M$ is weakly closed, namely $\M$ is a von Neumann algebra. 

Thus {\it observables} are selfadjoint elements $T$ of $\M$, {\it states} are normalised, normal, positive linear functionals $\f$, 
\[
\f(T) = \text{expected value of the observable $T$ in the state $\f$\ .}
\]
We regard $\M$ as a \textit{noncommutative space}. More precisely, $\M$ is the dual of a noncommutative measure space since, 
\[
\centerline{
$\M$ abelian von Neumann algebra $\Leftrightarrow$ $\M = L^\infty(X,\mu)$}
\]
namely $\M$ is isomorphic to $L^\infty(X,\mu)$
for some measure space $(X,\mu)$.

\subsection{Thermal equilibrium states}
 A primary role in statistical mechanics is played by the equilibrium distribution, that is classically given by the Gibbs distribution.

A {\it finite quantum system} is given by a matrix algebra $\gA$, the Hamiltonian $H$ is a positive selfadjoint element of $\gA$ and the time evolution is $\tau_t = {\rm Ad}e^{itH}$. An equilibrium state $\f$ at inverse temperature $\b$ is characterised by
the Gibbs condition that here is given by
\[
\f(X) = \frac{{\rm Tr}(e^{-\b H}X)}{{\rm Tr}(e^{-\b H})} \ .
\]
{\it What are the equilibrium states at infinite volume where there is no trace {\rm Tr} and no inner Hamiltonian?}

Let $\gA$ be a $C^*$-algebra, $\tau$ a one-parameter automorphism group of $\gA$. 
A state $\f$ of $\gA$ is KMS at inverse temperature $\b>0$ if for every $X,Y\in\gA$ there exists a complex function 
$\ F_{XY}$, analytic on the strip $S_\b=\{0<\Im z <\b\}$, bounded and continuous on the closure of $S_\b$, such that for all $t\in \mathbb R$
\textit{\begin{itemize}
\item[$(a)$] \textit{$F_{XY}(t)= \f\big(X\tau_t(Y)\big)$}
\item[$(b)$] \textit{$F_{XY}(t+i\b)=\f\big(\tau_t(Y)X\big)$}
\end{itemize}}
\vskip-0.3cm
\[
\begin{tabular}{c}
\centerline{\tiny  $\f\big(\tau_t(Y)X\big)$}\\
\hline
\centerline{$\big|${\tiny$\b$}}
 \\
\hline
\centerline{\tiny  ${\f\big(X\tau_t(Y)\big)}$}
\end{tabular}
\]
{\it The KMS states are the equilibrium states}. The  KMS and the Gibbs condition coincide when the system is finite. (See \cite{BR}). 

\subsection{Tomita-Takesaki modular theory}
Let $\M$ be a von Neumann algebra on a Hilbert space $\H$,  $\f = (\Om, \cdot\Om)$ the normal faithful state on $\M$ given by a cyclic and separating unit vector $\xi\in\H$. Then $\M$ linearly 
embeds into $\H$ by evaluating an operator $T\in\M$ on $\xi$; the embedding identifies, as linear spaces, $\M$ with a dense linear subspace of $\H$, so the $^*$-operation on $\M$ gives a densely defined, anti-linear operator $S_0$ on $\H$  
\[
\CD
\M @> X\mapsto X^* >\text{isometric} >\M
\\   @ V X\to X\Om V  V   @ V V X\to X\Om V \\  
\H @ > S_0:X\Om\mapsto X^*\Om >\text{non isometric}> \H
\endCD
\]
$S_0$ is a closable operator; let 
$S = \bar S_0$ be its closure. Then  $\Delta=S^*S > 0$ a positive selfadjoint operator on $\H$, the {\it modular operator}; $\log\Delta$ implements a canonical one-parameter automorphism group $\s$ of $\M$:
\[
s\in\mathbb R \mapsto \s_s^\f\in {\rm Aut}(\M)
\]
\[\s_s^\f(X) = \Delta^{is}X\Delta^{-is}
\]
the {\it modular automorphisms}.  So we have  an {\it intrinsic dynamics  associated with $\f$}. The $s$-parameter is 
the {\it modular time}. 

Moreover, if $S = J\Delta^{1/2}$ is the polar decomposition of $S$ we have a {\it canonical anti-isomorphism} between $\M$ and $\M'$  on $\H$
\[
J \M J = \M' \ .
\]
Some properties of the modular evolution are:

\medskip

$\bullet$ $\s^\f$ is a \textit{purely noncommutative} object (trivial in the commutative case). 
\medskip

$\bullet$ $\s^\f$ is a KMS \textit{thermal equilibrium evolution}: If $\f(X) = \text{Tr}(\r X)$ (type I case) then $\s_s^\f(X) = \r^{is}X\r^{-is}$.
The KMS inverse temperature is $\b = -1$. 
\medskip

$\bullet$ The \textit{arrow of modular time has a thermodynamical nature} as is associated with the sign of the inverse KMS temperature. 
\medskip

$\bullet$  The \textit{modular time is intrinsic modulo a scaling} $s\mapsto \s^\f_{-s/\b}$ that gives the inverse temperature $\b$. 
\medskip

$\bullet$ $\s^\f$ \textit{depends on the state $\f$ only up to inner automorphisms} (Connes' Radon-Nikodym theorem) \cite{C73}. 

\subsection{Time as thermodynamical effect}
By the above considerations, \textit{
if time is the modular time, then the time arrow is associated both with positive entropy and with quantum structure.} Here is a diagram that summarises our discussion so far:
\smallskip

\centerline{quantum physics} 
\centerline{$\updownarrow$}
\centerline{ KMS with positive temperature}
\centerline{
$\updownarrow$}
\centerline{modular time arrow}
\centerline{
$\updownarrow$}
\centerline{
 positive entropy}
\centerline{
$\updownarrow$}
\centerline{
thermodynamical arrow}
\bigskip

\noindent
We refer to \cite{Sew, CR, MR, BCL, Rov, Ron} for related discussions.
\subsection{Jones index}
Factors (von Neumann algebras with trivial center) are usually infinite-dimensional. For an inclusion of factors $\N\subset\M$ the Jones index $[\M : \N]$ measures the relative size of $\N$ in $\M$ of these infinite-dimensional objects. V. Jones showed \cite{Jo} a surprising result, the index values are quantised: 
\textit
{\[
[\M : \N] =  4\cos^2\!\left(\frac{\pi}{n}\right)\ ,\quad n=3, 4,\dots \quad {\rm or} \quad [\M : \N]\geq 4 \ .
\]}
Jones index appears in many places both in mathematics and in physics. See \cite{K86, Hi, L89, L90, L92} for the index of an inclusion of  factors of infinite type and \cite{GL} and refs. therein for the the non-trivial center case. 
\subsection{Local Quantum Field Theory}

In Quantum Field Theory, we have a quantum system with infinitely many degrees of freedom. The system is relativistic and particle creation and annihilation may occur. 
\[
\text{No mathematically rigorous QFT model with interaction still exists in 3+1 dimensions. }
\]
According to R. Haag \cite{Haag}, the physical content of QFT is to be encoded in the net of local von Neumann algebras: 
\[
O\ \text{spacetime regions} \mapsto \text{von Neumann algebras}\ {\mathcal A}(O),
\]
so to each region one associates the ``noncommutative functions" with support in $O$.
\medskip

A \textit{local net $\A$}  on the Minkowski spacetime $M$ is a map $O\subset M \mapsto  \A(O)\subset B(\H)$ such that associates with each spacetime region $O$ a von Neumann algebra $\A(O)$ on a fixed Hilbert space $\H$ with first principle physical requirements
(Haag-Kastler axioms, see \cite{Araki}):
\medskip

$\bullet$ \textit{\it Isotony}. $O_{1}\subset O_{2}\implies\A(O_{1})\subset\A(O_2)$  
\medskip

$\bullet$ \textit{\it Locality}.  $O_1$,
$O_2$ spacelike $\implies[\A(O_{1}),\A(O_{2})]=\nolinebreak\{0\}$ 
\medskip

$\bullet$  \textit{\it Poincar\'e  covariance}. There exists a unitary representation $U$ of the Poincar\'e $\P_+^\uparrow$ group such that $U(g)\A(O)U(g)^* = \A(gO)$, $g\in \P_+^\uparrow$ 
(in some context, the symmetry group could be the conformal group or other) 

\medskip

$\bullet$  \textit{\it Positive energy and vacuum vector}.  The spectrum of the restriction of $U$ to the translation group is contained in the forward light cone. There is a unique (up to a phase) $U$-invariant vector $\Omega\in\H$, and $\Omega$ is cyclic for the algebra generated by all the local von Neumann algebras $\A(O)$: 
\[
\text{$O\mapsto \A(O)$ may be viewed as a ``noncommutative chart'' }.
\]

\subsection{The Bisognano-Wichmann theorem.  Sewell's comment}
\label{BW}

By the Reeh-Schlieder theorem, the vacuum vector is cyclic and separating for the von Neumann algebra $\A(O)$ if both $O$ and its spacelike complement $O'$ have non-empty interior. 

Let $W$ be a wedge region of the Minkowski spacetime (Rindler spacetime) say  $W$ is the region $x_1 > |t|$. 
Under natural assumptions, the Bisognano-Wichmann theorem gives the vacuum modular operator and conjugation of $\A(W)$, in particular 
\[
\Delta^{-is}_W = U(L_W(2\pi s))\ ,
\]
where $L_W$ is the boost one-parameter group (pure Lorentz transformations) in the $x_1$ direction. 
\begin{figure}[h]  \begin{center}   {\footnotesize                
\begin{tikzpicture}[scale=0.8]
\draw[->, very thin] (-1,0)--(4,0) ;
\draw[->,very thin] (0,-2.4)--(0,3) ;
\draw [-] (0,0) -- (2.4,-2.4); 
\draw [-] (0,0) -- (2.8,2.8); 
\draw [dashed, ->, thick, domain= 0:1.72] plot ({cosh(\x)}, {sinh(\x)}); 
\draw [dashed, thick, domain= 0:1.62] plot ({cosh(\x)}, {-sinh(\x)}); 
\draw (1,0) node {$\bullet$};
\draw (0.2,3.0) node {$t$};
\draw (1.5,-0.25) node{ $1/a$};
\draw (4.1,-0.25) node{ $x_1$};
\draw (4.96,1.0) node {$t = a^{-1}\sinh 2\pi s $, $x_1 =a^{-1}\cosh 2\pi s$};
\draw (5.26,-1.0) node {trajectory unif. accelerated observer $\cal O$};
\end{tikzpicture}}\end{center}
\end{figure}
Sewell \cite{Sew} proposed in this context (or more generally as in Schwarzschild spacetime case) a model independent
derivation of the   Hawking-Unruh temperature \cite{Haw} \cite{Unr}, by means of the
Bisognano-Wichmann theorem.
The evolution
\[
\left\{ \aligned & t(s) =  a^{-1}\sinh s \\
& x_1(s) =  a^{-1}\cosh s
\endaligned \right.
\]
corresponds to an observer $\cal O$ moving within $W$ with uniform
acceleration $a$, and his proper time is equal to $s/a$. $W$
is a natural horizon for this observer, since he cannot send a
signal out of $W$ and receive it back. The von Neumann algebra
$\A(W)$ is therefore the proper (global) observable algebra for such a
mover. The
 proper time translations $\cal O$ are  given by the 
one-parameter automorphism group $\a_{as} = {\rm Ad}U(L_W(a s))$ of $\A(W)$ corresponding
to the rescaled pure Lorentz transformations leaving $W$ invariant.
By the Bisognano-Wichmann theorem $\a_{as}$ satisfies the KMS condition at
inverse temperature 
\[
\b = \frac{2\pi}{a}
\]
with respect to  the restriction of
the vacuum state $\f$ to  $\A(W)$, i.e. the latter is a 
thermal equilibrium state. By  Einstein's equivalence
principle, one can identify $W$ with the outside region of the black
hole and interpret the thermal outcome as a gravitational (black
hole) effect, so:
\[
\text{time is a geodesic flow, a quantum gravitational effect,} 
\]
$\b = 2\pi/a$ is indeed the Hawking-Unruh temperature. Cf. \cite{CR, MR}. 

The geometric expression of the modular group associated with a bounded region in conformal QFT is given in \cite{HL}. We refer to \cite{DM} and refs. therein for a discussion of the Bisognano-Wichmann property in a general context. 

\subsection{Representations. Index and statistics}
\label{Rep}

Doplicher-Haag-Roberts considered localised (DHR) \emph{representations} of a local net $\A$ \cite{DHR}. It turns out that a DHR representation is unitary equivalent to a localised endomorphism $\rho$ of a natural $C^*$-algebra generated by the net, one can define the statistics of $\r$ and the {\it statistical dimension} $d(\r)$ of $\r$. 

The {\it index-statistics relation} \cite{L89,L90} gives
$
d(\r) = \Big[ \A(O) : \r\big(\A(O)\big)\Big]^{\frac12}
$ if $\r$ is a DHR endomorphism localised in a causally complete region $O$. So
\[
{\text{DHR dimension} =\sqrt{\text{Jones index}}}
\]
and we have an equality between a physical index  
and an analytical index that provides a bridge between two subjects. See \cite{L89, LR,L01} for a conceptual discussion on the matter.

\section{Intrinsic bounds on entropy}

\subsection{Araki's relative entropy}

Quantum Information is providing an increasingly important interplay with Quantum Field Theory, that naturally originated in the framework of quantum black hole thermodynamics (see \cite{Wald}). The first non commutative entropy notion, von Neumann's quantum entropy, was originally designed as a Quantum Mechanical version of Shannon's entropy: if a state $\psi$ is given a density matrix $\r_\psi$
\[
\text{von Neumann entropy of $\psi$}: \ - {\rm Tr}(\r_\psi\log\r_\psi) \ .
\]
As said, an infinite quantum system is described by a von Neumann algebra $\M$ (e.g. a local von Neumann algebra $\A(O)$ in QFT) that is typically not of type $I$ (see \cite{L82}), no trace nor density matrix exists and von Neumann entropy is undefined. Nonetheless, the Tomita-Takesaki modular theory applies and one 
may considers  \cite{Ar}
\[
\text{Araki's relative entropy}:\  S(\f |\!| \psi) = -(\xi, \log\Delta_{\eta,\xi}\xi)
\]
between $\psi$ and the vacuum state $\f$, that extends Umegaki's type $I$ notion
\[
S(\f |\!|\psi) = {\rm Tr}\big(\r_\f(\log\r_\f - \log \r_\psi)\big)
\]
and is defined in general; here $\xi$ and $\eta$ are cyclic vector representatives of $\f$, $\psi$ and  $\Delta_{\eta,\xi}$ is the associated relative modular operator (see \cite{OP,Wit}). 
            
$S(\f |\!|\psi)$ measures how $\psi$ deviates from $\f$. 
From the information theoretical viewpoint, $S(\f |\!|\psi)$ is the mean value in the state $\f$ of the difference
between the information carried by the state $\psi$ and the state $\f$. 
Two fundamental properties of the relative entropy are its  \textit{positivity}
\[
S(\f |\!| \psi) \geq 0
\]
and \textit{monotonicity} w.r.t. a completely positive, normal unital map $\a$ (see below):
\[
S(\f\cdot\a|\!|  \psi\cdot\a) \leq  S(\f |\!|  \psi) \ .
\]
Relative entropy is one of the key concepts in information theory. \textit{We take the point of view that relative entropy is a primary concept and all other entropy notions are derived concepts}. 

\subsection{Analog of the Kac-Wakimoto formula}
The root of our work relies in this formula for the incremental free energy of a black hole given in \cite{L97}
(see also \cite{KL05}).
Let $H_\r$ be the Hamiltonian for a uniformly accelerated observer in the Minkowski spacetime with acceleration $a>0$ in a representation $\r$ localised in the wedge for $W$ (the generator of the rescaled boost unitary group in the representation $\r$, see Sections \ref{BW}, \ref{Rep}). We have
\textit{\ben
(\Om, e^{-s H_\r} \Om)\big\vert_{s= \b} = d(\r)
\een}\label{fKW}
with $\Om$ the vacuum vector, $\b = 2\pi/a$ the inverse Hawking-Unruh temperature and $d(\r)^2$  is Jones' index. 

The left hand side in \eqref{fKW} is a {\it generalised partition formula}, so $\log d(\r)$ has an \textit{entropy meaning}.  
(Cf. \cite{PP} for the entropy meaning of the logarithm of the index). 

\subsection{Completely positive maps, quantum channels and entropy}
Let $\N,\M$ be von Neumann algebras. A linear map $\a:\N\to \M$ is {\it completely positive} if 
\[
\a\otimes {\rm id}_n : \N\otimes {\rm Mat}_n(\mathbb C) \to \M\otimes {\rm Mat}_n(\mathbb C) 
\]
is positive for every positive integer $ n$. A completely positive map represents a {\it quantum operation}. We shall consider here only normal, unital completely positive maps. 

Let $\om$ be a  faithful normal state of $\M$ and  $\a: \N\to\M$ completely positive map as above. Set
\[
{\rm H}_\om (\a) \equiv \sup_{(\om_i)} \sum_{i}  S(\om |\!|  \om_i) - S(\om\cdot\a |\!| \om_i\cdot\a)
\]
with  supremum  over all finite families of positive linear functionals  $\om_i$  with $\sum_i \om_i = \om$ (one can easily define the relative entropy w.r.t. non-normalised states too. See \cite{CNT, Hi, OP} for other entropy aspects).

The \textit{conditional entropy} $H(\a)$ of $\a$ is defined by
\[
{\rm H}(\a) = \inf_\om {\rm H}_\om(\a)
\]
with infimum over all states $\om$ for $\a$ that are naturally represented by a cyclic vector (see \cite{LXbek}). Clearly \textit{${\rm H}(\a)\geq 0$} because 
${\rm H}_\om(\a)\geq 0$ by the monotonicity of the relative entropy. 

We shall say that $\a$ is a \textit{quantum channel} if its conditional entropy ${\rm H}(\a)$ is finite. 

\subsection{Bimodules and completely positive maps}
Let $\a : \N \to \M$ be a completely positive, normal, unital map and $\om$ a faithful normal state of $\M$.
There exists an $\N-\M$ {\it bimodule} $\H_\a$, namely a Hilbert space $\H_\a$ equipped 
with normal left/right actions $\ell_\a$/$r_\a$ of $\N$/$\M$ on $\H_\a$, with a cyclic vector $\xi_\a\in\H_\a$ for $\ell_\a(\N)\vee r_\a(\M)$, such that
\[
(\xi_\a, \ell_\a(n)\xi_\a) = \om_{\rm out}(n)\, ,\quad (\xi_\a ,r_\a(m)\xi_\a) = \om_{\rm in}(m) \, ,
\]
with $\om_{\rm in} \equiv\om$, $\om_{\rm out} \equiv \om_{\rm in}\cdot\a$, see \cite{C}. The converse is true (modulo unitary equivalence) so, given $\f$ we have a one-to-one correspondence.
\[
\text{completely positive  map}\ \a \longleftrightarrow \text{cyclic bimodule}\ \H_\a
\]
with the above properties. 
We have
\[
{H(\a) = \log \Ind(\a)}
\]
where $\Ind(\H)= \Ind(\H_\a)$ is the Jones' index (more precisely, the minimal index) of $\ell_\a(\N)\subset r_\a(\M)'$, \cite{LXbek}. 

\subsection{Generalisation of Stinespring dilation}
Let $\a: \N\to \M$ be a normal, completely positive, unital map between the von Neumann algebras $\N$ and $\M$. 
We shall say that a dilation pair  $(\r , v)$ with  $\r: \N\to \M$ a normal homomorphism and $v\in \M$ an isometry such that
\[
\a(n) = v^*\r(n) v\ , \quad n\in \N\ .
\]
$(\r, v)$ is a  {\it minimal} dilation pair for $\a$ if  the left support of 
$\r(\N)v\H$ is equal to 1. 

If $\N$, $\M$ are properly infinite (on a separable Hilbert space),  there exists a minimal dilation pair $(\r, v)$ for $\a$ and is unique: 
if $(\r_1, v_1)$ is another minimal  pair, there exists a unique unitary $u\in \M$ such that 
\[
u\r(n) = \r_1(n)u\, ,\quad v_1 = uv,\quad n\in\N\ .
\]
We have
\[
H(\a) = \log \Ind(\rho)\qquad \text{(minimal index)}. 
\]
\subsection{Promoting modular theory to the bimodule setting}

Let $\H$ be an $\N-\M$-bimodule with finite Jones' index $\Ind(\H)$. 
Given faithful, normal, states $\f,\psi$ on $\N$ and $\M$, we define the \textit{modular operator} $\Delta_\H(\f |\psi)$  of $\H$ with respect to $\f,\psi$ as
\textit{\ben
\Delta_\H(\f |\psi) \equiv d(\f\cdot\ell^{-1})\big/ d(\psi\cdot r^{-1}\cdot\e) \ ;
\een}\label{modH}
here $\Delta_\H(\f |\psi) \equiv d(\f\cdot\ell^{-1})\big/ d(\psi\cdot r^{-1}\cdot\e)$ is Connes' {\it spatial derivative} \cite{C80} between the states 
$\f\cdot\ell^{-1}$ on $\ell_\a(\N)$ and $\psi\cdot r^{-1}\cdot\e$ on  $\ell_\a(\N)'$
and  $\e: \ell(\N)' \to r(\M)$ is the {\it minimal conditional expectation} (see \cite{Hi, L90,Y}). 

\textit{$\log \Delta_\H(\f |\psi) $} is called the \textit{modular Hamiltonian} of the bimodule $\H$. 
The \textit{modular Hamiltonian of a quantum channel $\a$} 
is the modular Hamiltonian of the  bimodule  $\H_\a$ is associated with $\a$. 

\subsection{Properties of the modular Hamiltonian}
\label{Dprop}
For simplicity, let us assume here that $\N$, $\M$ are infinite factors and that our bimodules have finite index. 
The modular Hamiltonian satisfies the following natural properties \cite{L18L}:
\[
\Delta^{is}_\H(\f_1|\f_2)\ell(n)\Delta^{-is}_\H(\f_1|\f_2) = \ell\big(\s^{\f_1}_s(n)\big), \quad n\in\N,
\]
\[
\Delta^{it}_\H(\f_1|\f_2)r(m)\Delta^{-is}_\H(\f_1|\f_2)= r\big(\s^{\f_2}_{s}(m)\big), \quad m\in\M,
\]
(implements the dynamics)
\[
\Delta^{is}_\H(\f_1|\f_2) \otimes \Delta^{is}_{\K}(\f_2|\f_3) = \Delta_{\H\otimes\K}^{is}(\f_1|\f_3)
\]
(additivity of the energy)
\[
\Delta^{it}_{\bar\H}(\f_2|\f_1) = d_\H^{-i2t}\, \overline{\Delta^{it}_{\H}(\f_1|\f_2)}
\]
(charge symmetry, up to a phase). 

If $T:\H\to\H'$ is a bimodule intertwiner, then
\[
T\Delta^{it}_\H(\f_1|\f_2) = (d_{\H'}/d_\H)^{it} \Delta^{it}_{\H'}(\f_1|\f_2)T 
\]
(functoriality, up to a phase). 

Here above, the tensor product is Connes's {\it bimodule tensor product} \cite{C} w.r.t.  $\f_2$ and  $d_\H = \sqrt{\Ind(\H)}$. 

\subsection{Physical Hamiltonian of a finite-index bimodule}
In order to remove the appearing of a phase in the natural properties in Sect. \ref{Dprop}, 
we may here easily modify the modular Hamiltonian so to fulfil the right physical requirements (additivity of energy, symmetry under charge conjugation, etc.). The operator
\[
K(\f_1|\f_2)\equiv -\log \Delta_\H(\f_1|\f_2) - \log d
\]
is the \textit{physical Hamiltonian} (at inverse temperature 1). Here $d = d_{\H}$. 

The physical Hamiltonian at inverse temperature $\b >0$ is then given by
\[
\b^{-1}K(\f_1|\f_2) =
-\b^{-1}\log \Delta -\b^{-1}\log d
\]
with $\Delta = \Delta_\H(\f_1|\f_2)$. 

There are two steps in passing from the modular Hamiltonian to the physical Hamiltonian:
\[
-\log \Delta\ \xrightarrow{\rm shifting}\  -\log \Delta - \log d\ \xrightarrow{\rm scaling}\ \b^{-1}\big(-\log \Delta - \log d\big) \ .
\]
Notice that \textit{the shifting is \textit{intrinsic}, the scaling is to be determined by the context}.

The physical Hamiltonian then satisfies the natural properties as in Sect. \ref{Dprop}, without phase corrections. 

\subsection{Modular and Physical Hamiltonians of  a quantum channel}

We now are going to compare two states of physical systems; $\om_{\rm in}$ is a suitable initial state, e.g. the
vacuum in QFT, and $\om_{\rm out}$ is a state that is reached from $\om_{\rm in}$ by some physically realisable process (quantum channel).

Let $\a:\N\to\M$ be a quantum channel and $\om_{\rm in}$ a faithful normal state of $\M$. Set $\om_{\rm out} = \om_{\rm in}\cdot\a$. 
\[
 \Delta_\a \equiv  \Delta_{\H_\a}(\om_{\rm in}|\om_{\rm out})\ ,
\]
\[
K_\a = \b^{-1}\big(-\log \Delta_{\a} - \log d_{\H_\a}\big) \ .
\]
They are, respectively, the {\it modular and the physical Hamiltonian} at inverse temperature $\b$ of the quantum channel $\a$, relative to the state transfer with input state $\om_{\rm in}$.  

\subsection{Thermodynamical quantities}
We  define the \textit{entropy} $S\equiv S_{\a,\om_{\rm in}}$ of a quantum channel $\a$ with respect to the initial state $\om_{\rm in}$ by
\[
S = -(\hat\xi , \log \Delta_\a\hat\xi) \ ,
\]
where $\hat\xi$ is a vector representative of the state $\om_{\rm in} \cdot r^{-1}\cdot\e$ in $\H_\a$. 

The quantity 
\[
E_\a = (\hat\xi, K_\a\hat\xi)
\]
is the \textit{relative energy} with respect to the states $\om_{\rm in}$ and $\om_{\rm out}$. 

The  \textit{free energy} $F_\a$ is now defined by the relative partition function
\[
F_\a =   \b^{-1}\log(\hat\xi, e^{-\b K_\a}\hat\xi) \ .
\]
As in eq. \eqref{fKW} one gets
\ben\label{f2}
\log (\hat\xi, e^{-\b K_\a}\hat\xi) = \log d(\a) = \log \Ind(\a)^{1/2} = \frac12 H(\a) \ .
\een
Remarkably, $F_\a$ satisfies the \textit{thermodynamical relation} at temperature $ T = \b^{-1}$:
\ben\label{TR}
F_\a = E_\a - TS_\a \ .
\een

\section{Applications}

\subsection{Landauer's bound for infinite systems}
In short, Landauer's principle states that ``information is physical" \cite{Lan}. One of its main motivations was to provide a solution to the Maxwell's demon paradox. More extensively, it can be stated as follows:

\noindent
{\it Any logically irreversible manipulation of information, such as the erasure of a bit or the merging of two computation paths, must be accompanied by a corresponding entropy increase in non-information bearing degrees of freedom of the information processing apparatus or its environment} \cite{Ben}. 

Let $\a:\N \to \M$ be a quantum channel between quantum systems $\N$, $\M$. 
As a consequence of eq. \eqref{f2}, if $\a$ is irreversible, namely $\Ind(\a) > 1$, we have
\ben\label{Lb}
F_\a \geq \frac12 k T \log 2 \ ,
\een
where we have now inserted the Boltzmann constant $k$ (usually we use natural units), because the smallest index value greater than one is 2. (We are using the definition in \cite{LXbek}, rather than in \cite{Lan}, for the modular Hamiltonian, namely the conditional expectation is inserted in the right, rather than left, action term; this amounts to a change of sign, in particular for the free energy). 

The original {\it  Landauer's lower bound} is 
\[
F_\a \geq  k T \log 2 \ .
\] 
In the finite-dimensional context, the the smallest index value greater than one is 4, so our lower bound reduces to the original Landauer's bound. Yet, our bound holds in the infinite-dimensional context, in full generality. 

\subsection{ Bekenstein's bound}
The Bekenstein bound is a universal limit on the entropy that can be contained in a physical system of given size and total energy \cite{Bek}. 
If
a system of total energy $E$, including rest mass, is enclosed in a sphere of radius $R$, then the entropy $S$ of the system is bounded by
\[
S\leq \lambda RE\ ,
\]
where $\lambda >0$ is a constant (the value $\lambda = 2\pi$ is often proposed). 

In \cite{Ca}, H. Casini gave an interesting derivation for this bound, based on relative entropy considerations, although in the type $I$ case. 
In QFT, the relative entropy $S(\f |\!| \psi)$ subtracts indeed the entanglement ultraviolet divergences, common to two states $\f$ and $\psi$, and is finite for a dense set of states. A similar subtraction can be done to define a localised form of energy. 

Local von Neumann algebras $\A(O)$ in QFT are typically of type $III$ (see \cite{L82}) and no trace does exist. By the above explained methods, we have a rigorous, general derivation of the Bekenstein upper limit for the entropy/information.

Let $\a: \N \to \M$ be a quantum channel between the factors $\N$ and $\M$. 
As we have $F = \frac12 \b^{-1}{\rm H}(\a)$ and  $H(\a) \geq 0$  (monotonicity of the entropy), then
\[
{F \geq 0} \quad \text{(positivity of the free energy).}
\]
So the thermodynamical relation \eqref{TR}
$
F = E - \b^{-1} S
$
entails the general, rigorous version of the Bekenstein bound 
\[
 S   \leq \b E
\]
In particular, for a Schwarzschild black hole spacetime,
the vacuum modular group of $\A(O)$ is geodesic, with $O$
the black hole exterior region in the Schwarzschild-Kruskal spacetime.
The KMS temperature is the Hawking temperature 
$T = 1/8\pi M = 1/ 4\pi R$
with $R= 2M$ the Schwarzschild radius, thus 
\[
S \leq 4\pi R E 
\]
with $S = S_{\a,\f}$ the entropy associated with the state transfer of the vacuum state $\om$ by a quantum channel $\a$, and $E$ the corresponding relative energy in the region $O$.  

\section{The information in a wave packet and QNEC}
\subsection{Energy conditions in QFT}

As is known, positivity of the energy plays an important role in classical physics, general relativity in particular. 
In QFT, the local energy may have negative density states \cite{EGJ}, although energy lower bounds may occur (see \cite{ FH,Wie}). 

The \textit{Quantum Null Energy Condition (QNEC)} was considered in \cite{BFKLW}
(see  \cite{CF}). It is associated with a deformation $W_\l$ of a wedge $W$ in the null direction (see Figure 1 where the boundary deformation is given by the function $f_\l$, with $\l$ the deformation parameter).
By physical arguments, the QNEC reduces to the inequality
\[
S''(\lambda) \geq 0 \ .
\]
Here, for every normal faithful state $\psi$ on $\A(W_\l)$,  $S(\l)$ is Araki's relative entropy $S(\psi |\!| \f)$  between $\psi$ and the vacuum state $\f$ on $W_\l$ and $S''(\l)$ is the second derivative of $S(\l)$ w.r.t. $\l$.  
\begin{figure}[h]
\centering
\begin{tikzpicture} \label{deform}
\draw [->] (4,0) -- (5,2); 
\draw [->] (-4,0) -- (-3,2); 
\draw [->] (-4,0) -- (-3,-0.5);
\draw [->] (4,0) -- (5,-0.5); 
\draw [very thick, domain= 0:1] plot (\x/0.5 - 1, {\x- \x*\x*\x} ); 
\draw[ very thick] [-] (-4,-0) -- (-1,0); 
\draw[very thick] [-] (1,0) -- (4,0); 
\draw[ thin] [-] (-1,0) -- (1,0); 
\draw   (5.5,2) node {$u$};
 \draw (5.5,-0.5) node {$v$};
 \draw  (4.4,1.7) node {$W_\l$};
 \draw (4.2,-0.5) node {$W_\l$};
 \draw (0.06,0.1) node {$f$};
 \draw (0.34,1.2) node {$f_\l$};
 \draw [very thick, dashed, domain= 0:1] plot (\x/0.5 - 1, {1.6*\x- 1.6*\x*\x*\x} ); 
 \draw [very thick, dashed, domain= 0:1] plot (\x/0.5 - 1, {2.2*\x- 2.2*\x*\x*\x} ); 
\draw [-][dotted,thick] (-1,0) -- (-0.5,-0.25);
\draw [-][dotted,thick] (1,0) -- (1.5,-0.25);
\end{tikzpicture}
\caption{\footnotesize The function $f$ is the boundary of the deformed region on the null horizon. }
\end{figure}
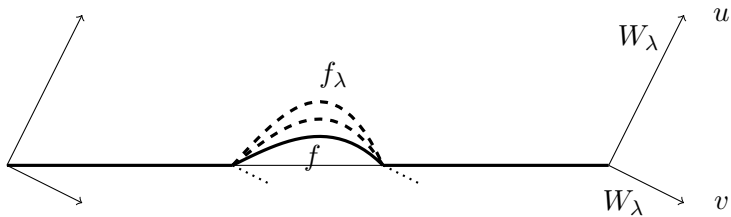
The positivity of the second derivative of the relative entropy is surprising as it does not hold for for arbitrary systems. 

In the following, we shall consider the particular, constant translation case, which is however the general case in the two-dimensional Minkowski spacetime context. 

\subsubsection{Entropy of coherent states: U(1)-current model}
Let's consider the case of the $U(1)$-current $j$ on the real line and the associated local conformal net $\A$ of von Neumann algebras on $\mathbb R$. A localised endomorphism of $\A$ is associated with a real function $\ell$ in $S(\mathbb R)$, its unitary equivalence class is labeled by the charge $\int\ell$, see \cite{BMT}. We have
\[
S(\l)  = \pi\!\int_{\l}^{+\infty} (x-\l)\ell^2(x){\rm d}x\ ,
\]
with $S(\l)$ vacuum relative entropy on $\A(\l, \infty)$ of the coherent state given by the vacuum excitation by the net automorphism induced by the shift of the current $j \mapsto j + \ell$, so
\[
S'(\l) = -\pi\!\int_\l^{+\infty} \ell^2(x){\rm d}x \leq 0\ ,
\]
\[
S''(\l) = \pi \ell^2(\l) \geq 0 
\]
and we have the 
\textit{positivity of $S''(\l)$} \cite{L19,CLR}. 

We refer to \cite{LXrel} for other aspects about entropy in Conformal QFT. 
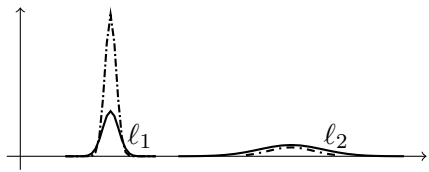
\begin{figure}[h]
\centering
\begin{tikzpicture}[scale = 0.6]
\draw [->] (-0.3,0) -- (9,0); 
\draw [->] (0,-0.3) -- (0,3.3); 
\draw [ thick, domain=-1:1] plot (\x + 2 , exp{-\x*\x*16}); 
\draw [ thick, densely dashdotted, domain=-1:1] plot (\x + 2 , pi*exp{-\x*\x*32}); 
\draw [ thick, domain=-2.5 :2.5] plot (\x +6, 1/4 *exp{-\x*\x}); 
\draw [ thick, dashdotted, domain=-1:1] plot (\x + 6 , 1/16 *pi*exp{-\x*\x*2});
\draw (2.65,0.45) node { $\ell_1$};
\draw (7,0.45) node { $\ell_2$};
\end{tikzpicture}
\caption{\footnotesize Two distributions (continuous lines) for the same charge $\int\ell_1 = \int\ell_2$. The dashed lines plot the corresponding entropy density rate $S''(\l)$: left: high entropy, right: low entropy. }
\end{figure}

\subsection{ The entropy of a vector relative to a real linear subspace}
Let $\H$ be a complex Hilbert space and $H\subset \H$ a closed, real linear subspace. 
A \textit{standard subspace} $H$ of $\H$ is a closed, real
linear subspace of $\H$ which is both \textit{cyclic} $(\overline{H +iH} =\H)$ 
and \textit{separating} $(H \cap iH = \{0\})$. $H$ is {\it standard}
iff the \textit{symplectic complement} $H'=\{\xi\in H : \Im(h,k) = 0\ \forall k\in H\}$ is standard.

Let $H$ be a standard subspace; the anti-linear operator 
$S : D(S) \subset \H \to \H$, 
\[
S:h + ik\to h - ik, \quad h,k \in H,
\]
is involutive, closed and densely defined, indeed
$S^*_H = S_{H'}$.
Let $S = J\Delta^{1/2}$ be the polar decomposition of $S=S_H$. 
Then $J$ is an anti-unitary involution, $\Delta > 0$  a non-singular, positive selfadjoint operator and
\[
\Delta^{it}H=H,\quad J H = H' \ .
\]
(one particle Tomita-Takesaki theorem), see \cite{L, LN}. 
In the following, we may assume that our real linear subspace $H$ is standard and factorial, namely $1$ is not an eigenvalue of $\Delta$. 

 Our analysis relies on the concept of entropy $S_k$ of a vector $k$ in a Hilbert space $\H$ with respect to the standard subspace $H$ of $\H$.
Our formula for the entropy for $S_k$ is
\[
S_k = \Im(k, P_H i \log\Delta\, k) \ .
\]
Here $P_H$ is the \textit{cutting projection} 
$P_H : H + H' \to H$,
\[
P_H: h + h' \mapsto h \ .
\]
In terms of the modular data,  $P_H$ is given by
\[
P_H = \Delta^{-1/2}(\Delta^{-1/2} - \Delta^{1/2})^{-1} 
+ J (\Delta^{-1/2} 
- \Delta^{1/2})^{-1} \ .
\]
\subsubsection{The information carried by a classical wave}
{\it Suppose that some information is encoded and transmitted by an undulatory signal, 
what is the information carried by the wave packet  in a given region at any given time?}

By a \textit{wave}, we shall here mean a solution of the Klein-Gordon equation
\[
(\square + m^2)\Phi = 0 \ ;
\]
we consider real waves  $\Phi$ 
with compactly supported, smooth time zero Cauchy data. 

Waves' linear spacetime is equipped with an independent symplectic form given by
$\frac12\int_{t = {\rm const}}(\Phi'\Psi - \Psi'\Phi )dx$ (here the dash denotes the time derivative), which is the imaginary part of a natural complex scalar product (depending on the mass $m$); so the wave linear space is dense in a Hilbert space $\H$. 

Waves with Cauchy data supported in the half-space $x^1 \geq 0$ form a real linear subspace $H(W)\subset\H$ ($W$ wedge), whose closure is standard. Then we define the entropy $S_\Phi$ of $\Phi$ as the entropy of the vector $\Phi$ with respect to the closure of $H(W)$ as above \cite{L19, CLR}. 

The entropy $S_{\Phi}(\l)$ of  $\Phi$ w.r.t. the wedge region $W_\l$ is given by 
\ben\label{Sp}
S_{\Phi}(\l) = 2\pi\int_{t = \l,\,  x^1\geq \l}(x^1 -\l)T_{00}(t,x) dx \ ,
\een
with $ T_{00} = \frac12 \big( \sum_{\mu=0}^d (\partial_\mu \Phi)^2 + m^2\Phi^2   \big)$ the classical energy-density of $\Phi$. 

In particular, the inequality
\ben\label{S2}
 S''_{\Phi}(\l)  \geq 0
\een
holds true.  

\subsection{QNEC for coherent states and constant translations}
Let $\A(W)$ be the von Neumann algebra associated with the Rindler wedge by the free, neutral quantum field theory.  
Given a real Klein-Gordon wave $\Phi$, consider the automorphism 
\[
\beta_\Phi = {\rm Ad}V(\Phi)^*|_{\A(W)} 
\]
and the vacuum entropy of $\beta_\Phi$ relative to the subwedge $W_\l = W + (\l,\l, 0,\dots,0)$, namely $\b_\Phi$ is the automorphism of $\A(W_\l)$ implemented by the Weyl unitary $V(\Phi)$ associated with $\Phi$.  

We have that the entropy of $\beta_\Phi$ with respect to the vacuum on $W_\l$ is given by
\[
S(\f_\Phi|_{\A(W_\l)} |\!| \f|_{\A(W_\l)})   =  S_\Phi(\l) \ ,
\]
where $\f$ is the vacuum state, $\f_\Phi = \f\cdot\b_\Phi$ and
$S_\Phi(\l)$ is the entropy of the wave $\Phi$ with respect to $H(W_\l)$ given in \eqref{Sp}. 

In particular, the positivity of $S''_\Phi(\l)$ in \eqref{S2} gives
\textit{the QNEC inequality 
for coherent states and constant null translations} for the free scalar quantum field case \cite{CLR}. 
We refer to \cite{CLR} for more details. 

We end up by summarising our construction by the following diagram:
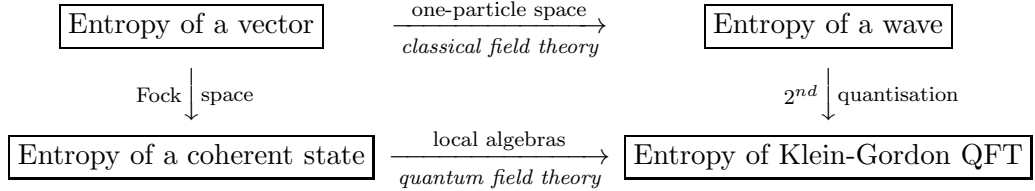
\[
\CD
\boxed{\text {Entropy of a vector}}    @>\text{one-particle space}> \textit{classical field theory}>  \boxed{ \text{Entropy of a wave}} \\ @V\text{Fock}V 
\text{space}\phantom{\bar X} V	@V\text{$2^{nd}$}V \text{quantisation}\phantom{\bar X}  V  \\  \boxed{\text{Entropy of a coherent state} }   @>\text{local algebras}> \textit{quantum field theory}>     \boxed{\text {Entropy of Klein-Gordon QFT}}\endCD
\]
So, {\it the entropy of a vector $\Phi$ with respect to a real linear subspace $H(W)$ has a double physical interpretation}: classically, it measures the information carried by a wave packet in the spacetime region $W$; from the quantum point of view, it gives the vacuum relative entropy, 
on the algebra $\A(W)$ associated with  $W$, of the coherent state induced by $\Phi$ on the Fock space.
\section{Final comments} The above thermodynamical interpretation of time concerns a system in thermal equilibrium; this is an asymptotic situation, that may apply to our present universe, not soon after the big bang. So far, the understanding of a system out of equilibrium is definitely incomplete and mathematical methods are to be developed, in particular Operator Algebraic ones.
However, Operator Algebras have recently provided a model independent framework, with effective computational methods, for non-equilibrium thermodynamics in low dimensional conformal field theory \cite{HL}. 

We finally mention that the Tomita-Takesaki modular theory also gives a powerful, surprising way to generate hidden symmetries in QFT, see \cite{Bor} and refs. therein.
\bigskip

\noindent
{\bf Acknowledgements.} 
Our discussion on time was initiated to be presented at the workshop ``The Origins and Evolution of Spacetime'' in the Vatican City at the Pontifical Lateran University, on November 27--28, 2018. We thank the IRAFS Director Gianfranco Basti and vice-Director Flavia Marcacci for the kind invitation. 
\smallskip

\noindent
We acknowledge the MIUR Excellence Department Project awarded to the Department of Mathematics, University of Rome Tor Vergata, CUP E83C18000100006.

\end{document}